\documentclass{aa}
\usepackage{graphics}

\begin{document}
\thesaurus{08(08.01.1; 08.16.3; 08.09.2 HD 84937;08.09.2 BD +362165;
08.09.2 BD +422667; 12.05.1}

\title{New high S/N observations of the $\mathrm{^6Li}$/$^7$Li 
blend\\
in \object{HD 84937} and  two other
metal-poor stars \thanks{ based on observations made 
 at the Canada-France-Hawaii Telescope and at Observatoire de 
Haute Provence}}

\author{R. Cayrel\inst{1} \and M. Spite\inst{2} \and F. Spite\inst{2} 
\and E. Vangioni-Flam\inst{3} \and M. Cass\'e\inst{3,4} \and J. 
Audouze\inst{3}}

\offprints{M. Spite}

\institute{ Observatoire de Paris, 
61 avenue de l'Observatoire, Paris, France \and 
Observatoire de Paris 
-Meudon, 92125 Meudon Cedex, France \and 
Institut d'Astrophysique, 98bis bd Arago, Paris, France
\and DAPNIA, CEA-Saclay, F-91191 Gif-sur-Yvette Cedex, France}

\date{Received  22 July  1998 /Accepted 4 January 1999}

\authorrunning{R. Cayrel  et al.}

\titlerunning{ $^6$Li in HD 84937}

\maketitle

\begin{abstract}

 High signal to noise ratio spectra have been obtained
 with the GECKO spectrograph at CFHT, at a spectral resolution 
 of 100 000, for three metal-poor
 stars in order to obtain more accurate ab\-un\-dances of 
 the very fragile element $^6$Li.  For two newly observed stars, 
 \object{BD +42 2667} and \object{BD +36 2165} 
 it appears that the first  may have a detectable amount  of $^6$Li, 
 whereas
 no $^6$Li is found in the second one. The  S/N  ratio of only a few 
 hundreds 
 obtained for these two faint stars preclude however a 
 firm conclusion.  For the third star, the well known object 
 \object{HD84937}, a  very high S/N of  650 per pixel (over 1000 per 
 resolved
 spectral element) was obtained, yielding greatly improved accuracy 
 over previous determinations. A value
 of $^6$Li / $^7$Li = $0.052 \pm 0.019$  (one sigma)
 is obtained.  We also conclude that the no- $^6$Li assumption is
 ruled out at the 95 per cent level, even in the most permissive case, 
 when a variation of all the other free parameters (wavelength zero-point,
 continuum location, macroturbulent broadening, abundance of $^7$Li )
 is allowed .  
  
 The possibility that the $^6$Li feature is an artifact due to a once 
 suspected
 binarity of \object{HD 84937} is discussed, with the conclusion that this 
 assumption
 is ruled out by the extant data on the radial velocity 
 of the object.
 The $^6$Li abundance is compared with recent models of 
 formation of the light elements Li, Be and B. This comparison shows 
 that $^6$Li
 is either undepleted, or only moderately depleted in \object{HD 84937}, 
 from its initial value.
  Under the  assumption that the atmospheric depletion of $^6$Li and 
 $^7$Li in stars is by slow mixing with hot layers (underneath the
 convective zone), in which these elements can burn, we conclude that  
 the depletion of $^7$Li  by this mechanism in \object{HD 84937} is 
 less than 0.1 dex. 

 This new upper limit to the efficiency of the depletion of $^7$Li 
 by slow mixing burning, in 
 a star located on the Spite plateau, leads to a more secure
 estimation of the primordial abundance of $^7$Li. However, the
 effect of temperature  inhomogeneities in the convective zone,
 on the derived abundance of lithium still remains to be 
 accurately determined.
 
   
 \end{abstract}
 
\section{Introduction}
The $^6$Li isotope is  an element of considerable interest, 
but extremely difficult to
 observe, as a weak component of a blend involving the much 
stronger doublet of 
$^7$Li, and with an isotopic separation of only 0.16 \AA. So far, 
it has been
 unambiguously detected 
 in only two stars, HD 84937 (Smith et al. \cite{SLN93}, 
Hobbs \& Thorburn (\cite{HT94}), 
and HD 338529 = BD +26 3578 (Smith \cite{Smith96}; Smith et al.
\cite{Smith98}). It has been unsuccessfully searched in five other 
halo stars by Hobbs \& Thorburn
(\cite{HT97}) and in 8 more by Smith et al. (\cite{Smith98}).  
 We have decided to reobserve HD84937 (for which the 1 $\sigma$ error
bar on the abundance is one-half of the abundance found; the best 
case!) with
 a signal/noise higher than
 obtained so far, and two other metal-poor stars in which we hoped to 
 have a new detection of $^6$Li. 
 
 Let us recall the double interest of obtaining evidence for $^6$Li , 
 in halo stars.  
 $^6$Li is a rare and fragile isotope of spallative origin, as are 
$^9$Be, $^{10}$B 
 and $^{11}$B. It is destroyed in stellar interiors at a temperature of 
about
2$\times10^6$ K,
 lower than the temperature of 2.5$\times10^6$ K at which $^7$Li is 
also destroyed.
 
 The first interest of finding $^6$Li in a halo star lying on the 
$^7$Li Spite plateau, is
 the strong presumption that, if $^6$Li has survived, it is 
 unlikely that the less fragile element $^7$Li
 has been significantly depleted in the star. The $^7$Li abundance 
on the Spite plateau can then
 be taken as representing the cosmological abundance of $^7$Li , 
predicted by the standard
 Big Bang nucleosynthesis (Schramm \& Turner 1998).\\
 
 The second interest is to compare the $^6$Li abundance  
 observed, with the predictions of 
 recent theoretical models of spallation  (Duncan et al. 
\cite{DLL92}, Vangioni-Flam et al.
 \cite{VLC94}, \cite{VCOF96}, \cite{VFCR97}). These models have been 
 triggered by the discovery that     
 the other spallative nuclei Be and B have an abundance increasing 
 linearily with metallicity in metal-poor stars   
 (Boesgaard and King \cite{BK93}, Boesgaard \cite{B96}, 
  Duncan et al. \cite{DPR97}, Molaro et al. \cite{MBCP97}).
More specifically, Vangioni-Flam et al. (\cite{VFCR97}) have predicted 
the 
production ratios of $^6$Li, $^7$Li, Be, $^9$B, $^{11}$B by low-energy
accelerated nuclei from SNe II, impinging the overwhelmingly 
dominant species in 
the early interstellar matter : H, and $^4$He. Recently 
Lemoine et al. (\cite{lemoine}) have dicussed extensively the abundance 
found earlier
for $^6$Li in the light of these recent predictions. Let us recall 
that the
dominant processus of formation for $^6$Li is the $\alpha+\alpha$ 
reaction, now
believed to occur between $\alpha $ particles accelerated in SNe II 
ejecta
colliding with $ \alpha $ particles of the surrounding interstellar 
matter (ISM). This
sets the initial abundance of $^6$Li in the star formed from this ISM, 
and part
of this $^6$Li  may be subsequently burnt in the star, where we 
observe the
remaining fraction.  The initial $^6$Li  can be inferred from the 
predicted $^6$Li /Be
 ratio, and the observed Be  abundance (see also Molaro et al.
\cite{MBCP97}). The depletion of $^6$Li
 within the star, by the $^6$Li(p,$^3$He)$^4$He process,
 is obtained by comparing the observed abundance of $^6$Li in the 
 star to the estimated initial abundance. Models of lithium burning by 
 mixing 
 of the convective zone with deeper layers allow to translate the 
 estimated
 depletion level of $^6$Li into some maximum depletion level for 
$^7$Li in the 
same star (Cayrel et al. \cite{CLM99}).\\
For this approach it is of course vital to be sure that $^6$Li is 
present. 
It is why we have decided to observe HD 84937 with a higher S/N 
ratio, and to observe
two other  halo stars, not yet analysed for $^6$Li, BD +42 2667 
and BD +36 2165. Actually BD + 42 2667 was also analysed in parallel 
by Smith et al. (\cite{Smith98}), but we were not aware of the work at 
the time of our observations.   
   The observations and the method of data reduction are described 
in section 2.  Section 3 is the analysis of the data in 
terms of $^7$Li and $^6$Li respective abundances.  Section 4 is 
devoted to the case of HD 84937.  Section 5 gives the results for the 
two other stars.  Section 6 discusses the significance of our 
results, and section 7 summarizes our conclusions.

\section{Observations and data reduction}  Spectra of the 
three selected stars HD 84937, BD+42 2667 and BD+36 2165 were 
obtained 
with the spectrograph GECKO at the 3.6m CFHT telescope in Hawaii (the 
log-book of the observations is given in Table 1).  The  
resolving 
power of the spectrograph, measured on thorium lines, is $R=100 000$.  The 
spectra were centered at 6710 \AA, the region of the Li\,I resonance doublet. 
In order to check the width of the lines, a spectrum of HD 84937 was also 
obtained in the region of the stronger, well defined, calcium line 
at $\lambda =6162$ \AA.  The  detector was a 2048x2048 15$\mu$m  pixels 
CCD fabricated by Loral.  The nominal gain of this CCD is 2.3 e 
ADU$^{-1}$, the read out noise 5.3 e pixel$^{-1}$, totally negligible
on our well exposed spectra. The pixel width was $15\mu$m, 
corresponding to .02698 \AA  . 

All the spectra have been reduced with a semi-auto\-matic code specially 
developed at  Observatoire de Paris-Meudon (Spite  \cite{Spi}). It 
performs the optimal extraction of the 
spectrum, the flat fielding and the wavelength calibration from 
the comparison lamp spectrum.  The wavelength calibration was performed
with a argon-thorium lamp. The laboratory wevelengths were  
taken in Palmer \& Engelman (\cite{PE83}) for thorium, and in
Kaufmann~\&~Edlen (\cite{KE74}) for Argon.
The rms of a third order polynomial fit, corrected for the actual number
of freedhom, is  of the order of 0.003\AA . It must be noted that,
 as it is well known (and e. g.  it is alluded to by Hobbs \& Thorburn
\cite{HT97}), the collimation angle between the beam of the lamp 
and the stellar beam, makes that there is a possible zero point shift between 
the wave length scale of the stellar and lamp spectra, and this shift 
may change during the night. Moreover, as the exposure time
for the thorium lamp is about one hour, the calibration spectra were 
usually taken at the beginning and at the end of the night. 
 The thorium spectra are quite usable
for establishing a calibration curve, but their zero-point can be 
slightly shifted 
at the time of the stellar exposure. This is why we have determined 
the zero-point of the wavelength scale mostly 
from the position of the calcium $\lambda =6717$ \AA \ line,
taken in the same exposure as the lithium feature.
Although the calcium $\lambda = 6162.172 $\AA \ line (wavelength 
from Sugar \& Corliss \cite{SC82}) is better defined, it was not 
used  for wawelength zero-point determination, for the reason explained 
above.
 
Flat-fielding has been done using a quartz lamp and a rapidly 
rotating 
hot star.  Due to small fringes produced by the CCD in the spectral 
range of the lithium line, the hot star has been finally preferred 
for 
flat fielding.
 
The stellar broadening ($\approx $.15  \AA \ is clearly larger than 
the spectrograph 
resolution ($\approx $ .07 \AA), and thus the stellar profiles are 
dominated by intrinsic 
stellar broadening.

\begin{table}
\caption[3]{Log book of the observations}
\begin{flushleft}
\begin{tabular}{lclrrccllllll}
\noalign{\smallskip}
\hline
\noalign{\smallskip}
 HD/BD   & date    &      & exp. & central &  estim. \\
mag.     & of obs. & UT   & time & wavele. &   S/N   \\
         &         & beg. & mn   &   \AA   &         \\
\hline
\noalign{\smallskip}
 84937      &  970420  &   6h10    & 120   &   6712  & 390  \\
V=8.28      &  970420  &   8h20    &  90   &   6712  & 300  \\
            &  970421  &   5h30    & 120   &   6162  & 380  \\
            &  970421  &   8h10    & 120   &   6712  & 350  \\
\noalign{\smallskip} 
\hline
+42 2667    &  970420  &  11h30    & 240   &   6712  & 240  \\
 V=9.86     &  970421  &  10h50    & 250   &   6712  & 230  \\
            &  970422  &  10h30    & 270   &   6711  & 280  \\
\noalign{\smallskip} 
\hline
+36 2165    &  970422  &   5h45    & 240   &   6711  & 280  \\

 V=9.78     \\ 
\noalign{\smallskip} 
\hline
\end{tabular}
\end{flushleft}
\end{table}

\section{Analysis} 
The models used in the analysis of the stars have been interpolated 
in 
the grid defined by Edvardsson et al.  (\cite{Edva93}) computed with an 
updated version 
of the MARCS code of Gustafsson et al.  (\cite{Gus75}) with improved UV line 
blanketing (see also Edvardsson et al.   \cite{Edva94}).  The physical 
parameters of the models have been taken from the literature and are 
given in table 2.

\begin{table}
\caption[3]{Main parameters of the adopted model atmospheres }
\begin{flushleft}
\begin{tabular}{lcccc}
\noalign{\smallskip}
\hline
\noalign{\smallskip}
star     & Teff & log g & [Fe/H] & Ref \\
HD/BD    &   K  & (CGS) &  dex  &     \\
\hline
\noalign{\smallskip}
 HD 84937 &  6300  &  4.0    & -2.3   & Nissen et al. (1994) \\
+42 2667  &  6059  &  4.0    & -1.7   & Rebolo et al. (1988) \\
+36 2165  &  6350  &  4.8    & -1.2   & Axer et al.   (1994) 
\\            

\noalign{\smallskip} 
\hline
\end{tabular}
\end{flushleft}
\end{table}

 The effective temperatures are all consistent with the Alon\-so et al. 
(\cite{Alonso96}) scale,
 based on the Infrared Flux Method  (Blackwell \& Shallis 
\cite{Blackwell77}). The microturbulence 
 has been set to 1.5  km sec$^{-1}$, following Smith et al. (\cite{SLN93}). 
 The 
$^6$Li /$^7$Li  ratio is not sensitive to the exact temperature
 adopted, only the absolute values are affected by the uncertainty
 on the zero point of the effective temperature scale.  
    To estimate the $^6$Li / $^7$Li
    ratio, we have proceeded very much as done in the seminal 
 paper by Smith et al. (1993).  Having used the same  atomic 
 data for the lithium feature, we have considered as adjustable 
parameters, 
 in the fitting of a synthetic spectrum to an observed spectrum, the 
five parameters
 (i) placement of the continuum, (ii) and (iii)   
  abundances of the two elements $^6$Li and $^7$Li , (iv) 
macroscopic broadening 
 of the lines (encompassing instrumental profile, star rotation and 
 macroscopic motions in the atmosphere of the object), and (v) 
wavelength
  zero-point adjustment.  
 The profile  of the macroscopic broadening was assumed to be 
gaussian and 
 is defined by its full width at half maximum FWHM. The 
 permissible range of this last parameter was determined by the 
 observation of the two calcium lines at 6162 and 6717 \AA.

   Each time, the quality of the fit
 between 27  points of the synthetic versus observed data points
 was computed and shown graphically. The quality of the fit has been 
 quantified by its $\chi ^2$,  The parameters have been determined by 
the 
 Maximum-Likelihood approach, i.e. by try trying to minimize 
 the $\chi ^2$, in the permissible range of variation of the parameters.   
  
 In order to compute the $\chi ^2$, one must know the value of the 
 observational standard error on each of the data point.  When 
several spectra were available we have  co-added them. As we had a 
 spectral resolution higher than needed to separate the isotopic
 shift of 0.16 \AA \ between the $^6$Li and $^7$Li lines,  we have 
 also convolved the coadded spectra by a gaussian of FWHM = 1.66 
 pixel $ \approx$ 0.045 \AA . Before this convolution the S/N on each 
 data point was 650 for HD 84937 (estimated in segments of the 
 continuum, and consistent with the expected photon noise).  After 
 convolution it raised to 1020 (noise 
 filtering). This represents a significant improvement in S/N, 
without a
 damaging loss of resolution. One must realize however, that 
 the classical $\chi^2$ test is not directly applicable to the 
 filtered spectrum, because the filtering introduces a correlation
 between  data point values. So we have computed (see appendix)
 the applicable pseudo-$\chi^2$ probability function   for the 
quantity $X^2$:
  
  $$ X^2= \sum_{i=1}^n ({O_i-C_i \over \sigma_i})^2  $$
  
  where the $O_i$ have been smothed by the convolution performed.  
Following
  a widely used notation, the $C_i$ are the computed values of the 
  synthetic spectrum, and  $\sigma_i$ is the noise on the
  $i_{th}$ data point.
As it will be explained in detail further, this approach permits a 
more powerful 
discrimination of the weak signal produced by the $^6$Li feature, 
thanks
to the improvement in the S/N ratio.

    Because of its particular interest we discuss in the 
 following section, in more detail, the case of HD 84937.
 
  \section{HD 84937}
\subsection{Fitting the observations with a synthetic spectrum}

\begin{figure}
\resizebox{\hsize}{!}{\includegraphics{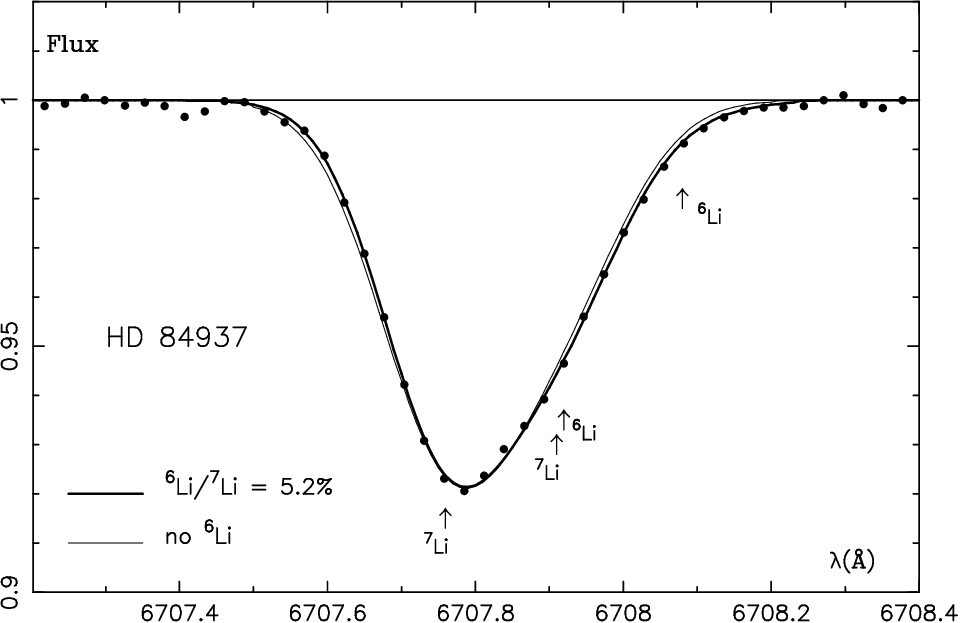}}
\caption{Observed 
  and synthetic profiles of the blend $^6$Li,$^7$Li for HD 84937.
  The synthetic profile has been computed using the stellar 
  parameters given in table 2.  
  Here the five parameters  
  (abundances of the two lithium isotopes, location of the 
  continuum, zero-point adjustment of wavelength scale, 
    FWHM of the 
  convolution taking care of the broadening of the lines by 
  macroscopic motions and  instrumental resolution),
   have been simulteanously least-square adjusted.  The dots 
  are the observations. The thick line represents the best fit 
 obtained 
  [log($^7$Li /H)=2.212, log($^6$Li /H)= 0.928, FWHM= 0.148 \AA , 
 wavelength shift = 10 m\AA , continuum redadjustment - 0.0006]. 
 The thin line represents the (degraded) fit when the $^6$Li abundance 
 is forced to zero, without further readjustment of the zero-point
 of the wave-length scale, but all the other parameters being readjusted: 
 $^6$Li /H=0.0, log($^7$Li /H) = 2.235, FWHM= 0.163 \AA ].} 
\label{figure1}
\end{figure}
\begin{figure}
 \resizebox{\hsize}{!}{\includegraphics{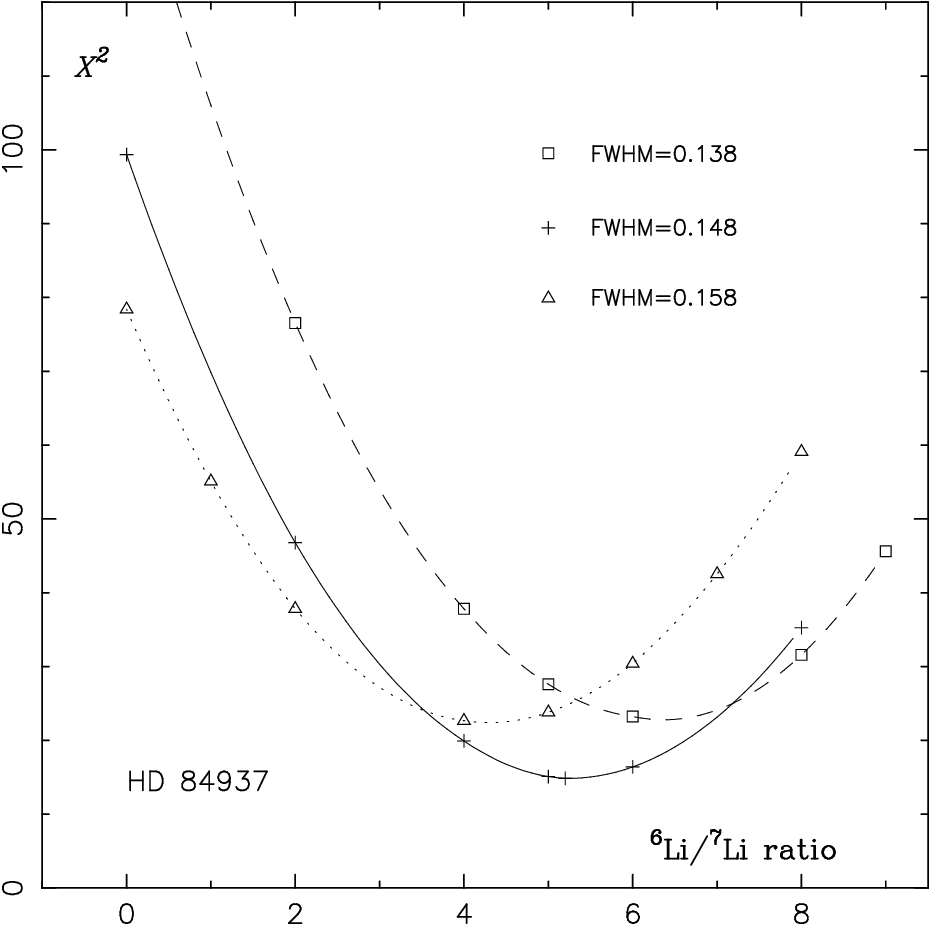}} \caption{ Fine 
 tuning 
  for the determination of the macroscopic broadening of the lithium 
 feature.
  A broadening of FWHM $\approx $ 0.15 gives the smallest residuals, 
 and this for
  a $^6$Li / $^7$Li ratio of 5.2 per cent. } 
\label{figure2}
\end{figure}

\begin{figure}
 \resizebox{\hsize}{!}{\includegraphics{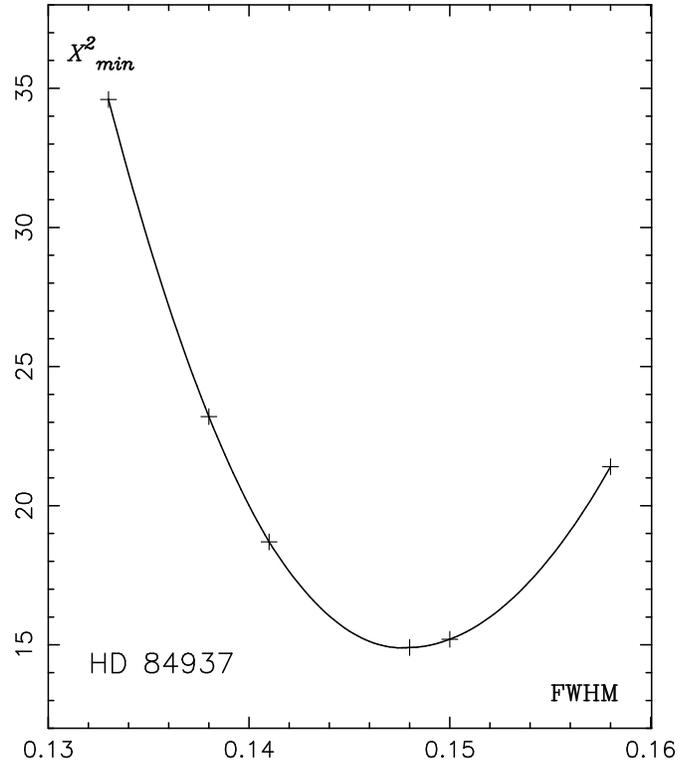}}
 \caption{ Interpolation, based on the results shown in fig. 2, for 
 finding
 the optimal value of the FWHM of the macroscopic broadening. The 
 value at the minimum minimorum is 0.148 \AA.}
 \label{figure3}
 \end{figure}
 
\begin{figure}
  \resizebox{\hsize}{!}{\includegraphics{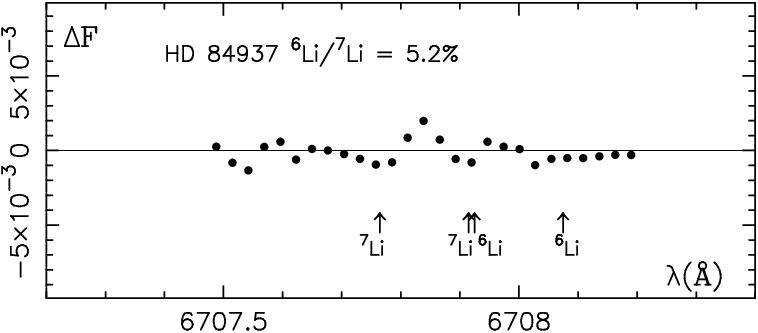}} 
\caption{Residuals (O-C) 
  of the best fit obtained with 5.2  per cent of $^6$Li .  The rms of 
  the fit between the synthetic spectrum  and the filtered observed 
 spectrum is 0.73E-03.
 The expected rms for  a perfect  synthetic model and the 
 noise level
 computed from photon noise is 0.98E-03 , so the fit is  
 slightly below the   
 expectation value, for a perfect model, quite possible for a 
 particular realization of a random noise. }
  \label{figure 4}
\end{figure}

\begin{figure}
\resizebox{\hsize}{!}{\includegraphics{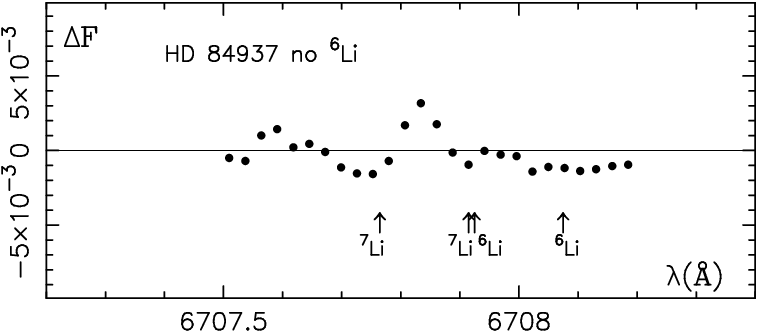}}
\caption{This
figure show the residuals of the best fit, when the abundance of $^6$Li 
is forced to zero,
all the four other parameters being allowed to vary.
This best fit for no $^6$Li has now a
rms of 1.19E-03,  
 significantly larger than the rms of the best fit with
 5.2 \%  of $^6$Li (0.73E-03). The $X^2$ of this fit is 39.8, 
 excluding the model
 at a level of confidence of 95.3 per cent.}     
  
  \label{figure5}
 \end{figure}
 
 \begin{figure}
 \resizebox{\hsize}{!}{\includegraphics{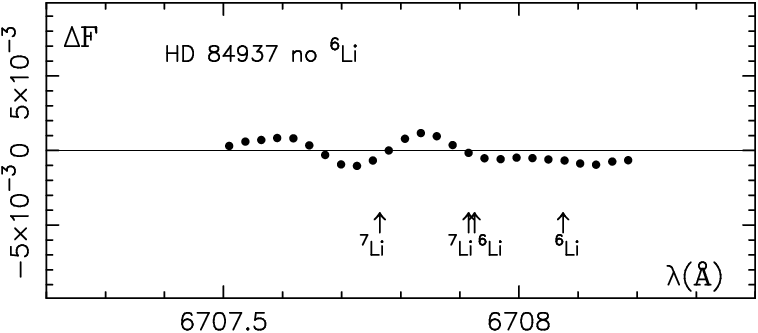}}
\caption{ Residual of the synthetic spectrum corresponding to the best
fit with no $^6$Li, and a wavelength offset of 0.005 \AA,  
relative to the synthetic spectrum 
corresponding to  the best fit with 5.2 \% of $^6$Li and no 
wavelength shift. These residuals are not produced by noise, but by 
the intrinsic
difference between the two spectra.
 The necessity of working at a  very high S/N ratio is demonstrated 
by this 
 comparison, the amplitude of this differential signal being of the 
order of a couple of 
 thousandths only. \emph{Note} the striking similarity with fig. 5, 
 giving the feeling that 
 the shape of the residuals of the fit with no $^6$Li shows 
 the signature of a wrong choice of the parameters.
 }
 
\label{figure6}
\end{figure}

It is often considered (Smith et al. \cite{SLN93}) 
that there are chiefly 
two methods for determining the isotopic ratio $^6$Li / $^7$Li. The 
presence
 of $^6$Li modifies the blend in (i) modifying the center of gravity 
 in wavelength of 
 the feature and (ii)  modifying the line profile by increasing the 
 opacity
 on the red wing. The first one is refered as the "c.o.g" method, 
 and the
 second as "line profile analysis". However both effects are combined 
 in the spectrum synthesis approach.  Smith et al. (\cite{SLN93})
 have pointed out that the  "c.o.g." method 
 is vulnerable to a  possible differential convective blue shift 
 between the formation
 of the Li blend and the only available other line in the spectral 
 field, the Ca I 6717.687 (Pierce and Breckinridge \cite{PB73}).  

 Actually, the blue shifts of solar lines have
 a fairly large range of values ( from -.7 km/s to zero, see  Allende 
 Prieto \& Garc\'{\i}a L\'opez \cite{APGL98}).
 Kurucz (\cite{K95}) has claimed that the effect of convective 
 temperature inhomogeneities could be
 considerably larger in metal-poor stars than in solar type 
 stars. Bonifacio\&Molaro (\cite{BM98}) have given a strong counter-argument concerning
 the claim of a large effect on the abundance of Li, but the question of a 
 wavelength shift remains.
 Furthermore, Smith et al. (\cite{Smith98}) (this work came to our
 knowledge after we wrote the submitted version of this paper)
 cite an unpublished work by Rosberg and Johansson giving a 
$\lambda$ of 6717.677 for the line instead of 6717.687 we first used. 
 They also mention that if the c.o.g method give a slight shift in 
 the right direction
 for HD 84937 it gives a blue shift, in contradiction with the 
 result  of the line profile
 analysis, for the other star, BD +26 3578.  Quantitatively speaking, 
 the c.o.g. method  is clearly far inferior to the line profile analysis, 
 a differential blue shift of only .001 \AA \ 
 corresponding to a relative change of 20 per cent  in the $^6$Li/ 
$^7$Li ratio (Smith et al. \cite{Smith98}),
 whereas such a shift has no influence on the line profile analysis, 
as being negligible
with respect to the isotopic shift of 0.16 \AA. Martin Asplund 
(private communication)
has investigated from hydrodynamical simulations of convection, the size of 
differential blue shifts between the lithium feature and Ca I, 6162 and 6717.
His preliminary results are that the Ca I lines are blue-shifted with respect 
to the lithium feature in HD 84937, by 0.2 to 0.4 km/s. This is enough to
completely kill the c.o.g. method, the corresponding variation in wavelength
being 9 m\AA  , larger than the shift between 5 per cent of $^6$Li 
and no $^6$Li at all.  
Therefore,  the safest approach, adopted here, is to consider that all relevant 
parameters are allowed
 to vary, including the zero-point of the wavelength scale, and then 
comparing this least-square solution to the values of the parameter
 which can be externally determined from other lines, {\em but 
allowing for the error bars on these 
 external determinations}. This is  what Smith et al. (\cite{SLN93}) 
 and Hobbs and Thorburn (\cite{HT97})
 have also done, concluding that the constraints on the 
$^6$Li/$^7$Li ratio
 was significantly relaxed by this approach, a price to pay for not 
being dependent on
assumptions on convective shifts, or ignoring that the macro-broadening of 
the lines cannot be fixed exactly.
   
  The situation is illustrated by the figures 1 to 4.  First we 
  consider as free parameters the position of the continuum, the 
abundance of the two isotopes 
  of lithium, the macroscopic broadening (due to stellar rotation, 
  macroscopic motions in the stellar atmosphere, and finite 
resolution 
  of the spectrograph) and the poorly determined zero-point of the 
wavelength scale.
The best fit to the data point, allowing the 5 parameters to 
vary, is shown by the full line of 
  fig.  1. This best fit is obtained for the following values :
  $^6$Li / $^7$Li = 5.2\%,
  log($^7$Li/H)=2.088, FWHM= 0.148 \AA, and an adjustment of the 
 ordinate scale
 of 0.06 per cent and of the zero point in wavelength by 10 m\AA .
       The residuals of this fit are shown in fig. 4.  The rms of the 
 fit on 27 points with the unfiltered data is 0.00120. The  
  expected value of the $\chi ^2$  for 27   points and 5 adjusted 
 parameters  
 is  22 for a perfect model, and the value found for the fit is 16.4 .
  With the filtered data the corresponding numbers are rms = 0.00073,
 22 and 15. 
   This means that  the model represents the data  without the need 
 of increasing the number of the parameters, and that
 the realization of the noise corresponding to our observation is 
 slightly better than its mathematical expectation,  but reasonably 
 close  
 to it. Now we can compare the value of the zero-point wavelength 
 shift and the value found for FWHM with the accuracy of our 
 wawelength
 scale and the value of FWHM determined from the Ca I lines. The 
 expected accuracy
 of our wavelength scale is dominated by the uncertainty on the 
 exact wawelength
 of the 6717 line, plus the unknown value of the differential blue-shift
 between the Li lines and the 6717 Ca I line. This is of the order 
 of 0.6 km/s, or 13 m\AA. The shift 
 obtained for the Li I feature {\em with the new value 
 6717.677  for the laboratory value of the Ca I line} is - 10 m\AA. This is
 within the uncertainties, although not in the direction predicted by
 Asplund. However, the Ca I line is strongly blended in the Sun, and very weak
 in HD 84937, so any conclusion based on such a small shift would be a 
 clear overinterpretation of the data.    
 The external values of FWHM are respectively 
 0.150 \AA  $\pm 0.01$ (6717 line)
 and 0.153 $\pm 0.08$ (6162 line, slightly saturated). The best fit 
 (0.148 \AA)  is totally compatible with the external values,
 which, unfortunately, do not allow to restrict efficiently the 
 internal uncertainties.

Now we retry the fit, forcing the abundance of $^6$Li to zero, 
but keeping the 4 other parameters free. We get a different 
solution,
with log($^7$Li/H) = 2.233, FWHM= 0.158, a differential redshift
with Ca I 6717.677 of 0.005 \AA \ and a negligible change in the 
continuum position. The residuals of this new fit are    
shown in fig. 5.   
  The rms of the fit is 0.00119 (with the continuum at 1.0) , corresponding 
to a value of $X^2$
of  39.8, rejecting this no-$^6$Li  model with a confidence level 
of 95.3 per cent. Would we have used the unfiltered spectrum,
the rms of the fit would have been 0.00155, corresponding to 
a $\chi^2$ of 27.3 for an expectation value of 27-4=23, which 
is excluded only at the 76 per cent level. However, many authors use a 
reduced relative $\chi^2$/$\nu $, not bothering for estimating
the noise independently. In that case the unfiltered $\chi^2$
being 16.4 for the best fit with 5.2 per cent of $^6$Li, one
can claim that the $\chi^2$ has been degraded in the ratio 
27.3/16.4 when the expected degradation is 24/23. The probability of 
such a difference
is only about 3 per cent. Of course the same approach can be also
applied to the filtered data, leading to an exclusion of 
the zero $^6$Li hyphothesis at the 99.5 per cent level.
We do not support these numbers, too optimistic, and we suggest 
to bring instead, now more attention to the run of the residuals
with wavelength.    
  
The shape of the residuals 
with wavelength is very suggestive of being not due to a 
random noise.
   In order to check if this is the case we have compared two 
synthetic
   spectra ({\em no noise here}), one corresponding to the absolute 
best fit,
the other one corresponding to the best fit, with no  $^6$Li , 
involving a 0.005 \AA \  shift in wavelength. The residuals of 
the second versus the first one are shown in fig. 6.
   The similarity between fig. 5 and 6 is striking. Fig. 6 also shows how 
important it
   is to reach a S/N of 1000, to  discriminate between these two 
cases. 
  We consider that the rejection of a zero-abundance of $^6$Li at the 
95 
  per cent level, {\em in the most permissive case}, represents a 
serious advance 
  with respect to   former works, done with a lesser S/N ratio.   
  
  A Bayesian approach of the problem, that we have not attempted, is 
likely to
  increase our value of $^6$Li , because of the {\em a priori} 
positive physical nature of the
  abundance of $^6$Li. For  shifts larger than 0.005 \AA \ the 
solution would 
  need $^6$Li in emission, not acceptable.
The 68 per cent probability around the optimal fit ($^6$Li /$^7$Li = 
0.052) can be computed when the continuum level,
and the two abundances are free parameters, because they occur 
linearily. We find
that this amounts  to $\pm$ .012 . An extra-allowance must be made 
for the impact of
the uncertainties on the other two parameters. For the FWHM of the 
broadening, if we accept
the full range of fig. 2 , with equal probability of the FWHM in the 
range, we obtain an
extra contribution of $\pm$ .007. The contribution of a zero-point 
shift in wavelength is
more difficult to estimate. A shift as large as the one needed for
 the no $^6$Li hyphothesis is 
largely rejected. We consider that
a shift of $\pm$ .002 is allowed. This  adds an extra-excursion of 
rms  0.013. The combined
effect is 0.019, adding quadratically . A Monte Carlo simulation 
would be more correct
to evaluate this uncertainty, but what is more important is to have 
excluded a zero abundance
of $^6$Li at the 95 per cent level.

 \subsection{Could HD 84937 be a binary star?}
 
 In one of their papers on proper motion stars (paper XII), Carney et 
 al. (\cite{Car94}) consider HD 84937 as a suspected spectroscopic binary.
  We have two recent measurements of
 this star with the OHP instrument ELODIE. We compare in table 3 
 these measurements with those of Carney and earlier.
 
 \begin{table}
 \caption[]{Radial velocities of HD 84937 from 1923}
 \begin{flushleft}
 \begin{tabular}{llll}
 \noalign{\smallskip}
 \hline
 \noalign{\smallskip}
 RV  & weight  & epoch & reference \\
 \noalign{\smallskip}
 \hline
 \noalign{\smallskip}
 -22.6 &  3 &   1923 &  ApJ 57, 149 \\
 -15.3  & undef & 1940 & PASP 52, 401 \\
 -12.9  & 2  & 1965 & MNRAS 129, 63 \\
 -16.4  & 9  & 1969 & AJ 74, 908 \\
 -13.1 & 8 &   1972 & AJ 77, 590 \\
 -16.7  &  8 & 1979    & IAUS 30, 57 \\
 -14.8  & 10 & 1988-94 & AJ 107, 2240  \\
 -14.95 & 20 & 19940305 & Mayor, unpubl.\\
 -15.17   & 20 & 19960415 & A\&A 338, 151 \\
 -14.8  &  15 & this paper & from Ca I 6162 \AA \\
 \noalign{\smallskip}
 \hline
 \end{tabular}
 \end{flushleft}
 \end{table}
     
 A rather arbitrary weight (20) has been attributed to the  two values obtained
 with ELODIE which have a standard error of 0.25 km sec$^{-1}$. There is no 
 indication of a systematic variation over 30 years. Moreover, Mayor
 (private communication 1998) indicates that the star, observed repeatedly 
 with CORAVEL during more than six thousand days, has no variation
 of radial velocity  at the level of 1 sigma = 1.19  km sec$^{-1}$.  
 If we imagine that the $^6$Li component is actually
 the $^7$Li component of a companion, the radial velocity 
difference 
 is 7 km sec$^{-1}$.
 In order to produce the 5 per cent  strength in the blend, the 
companion
 must be a dwarf of absolute magnitude about 6.8 having an effective 
temperature
 of 5400  K and a mass of 0.6 M$_{\sun}$, whereas the primary has a 
mass 
 of about 0.8 M$_{\sun }$.
 Because, $K_1$ and $K_2$ are the amplitudes of the radial velocity
 variation of the two components, we have the two constraints :
 $K_1+K_2  \geq 7$\ km sec$^{-1}$
 and :
 $$\sin^3 i \leq 1 $$
 and using the well known relation  Allen (\cite{AQ}):

 $$(M_1+M_2)\sin^3 i=1.035.10^{-7}(1-e^2)^{3/2}(K_1+K_2)^3P$$

 the following condition is obtained :

 $$P \leq 0.966.10^7(M_1+M_2)/(343(1-e^2)^{3/2}) $$

 Replacing $ (M_1+M_2) $ by its value and taking a typical value of 
0.3 for {\it e}, we get
 $$ P \leq 45000 \, \mathrm{days}$$ or 124 years.
 This condition is barely acceptable if there is no sign of variation 
of the
 radial velocity in 30 years, which represents one fourth of the 
maximum period.
 We therefore consider as rather unlikely the assumption that we see 
a second
 $^7$Li blend instead of a $^6$Li component.
 A second implication of the same assumption would be that not only 
 the lithium blend would be affected,
  but also all other absorption lines. We have computed the effect
 on the line Ca I 6162, observed in the same run. The computed and 
 observed profiles are strongly discrepant (see fig. 7).
 \begin{figure}
 \resizebox{\hsize}{!}{\includegraphics{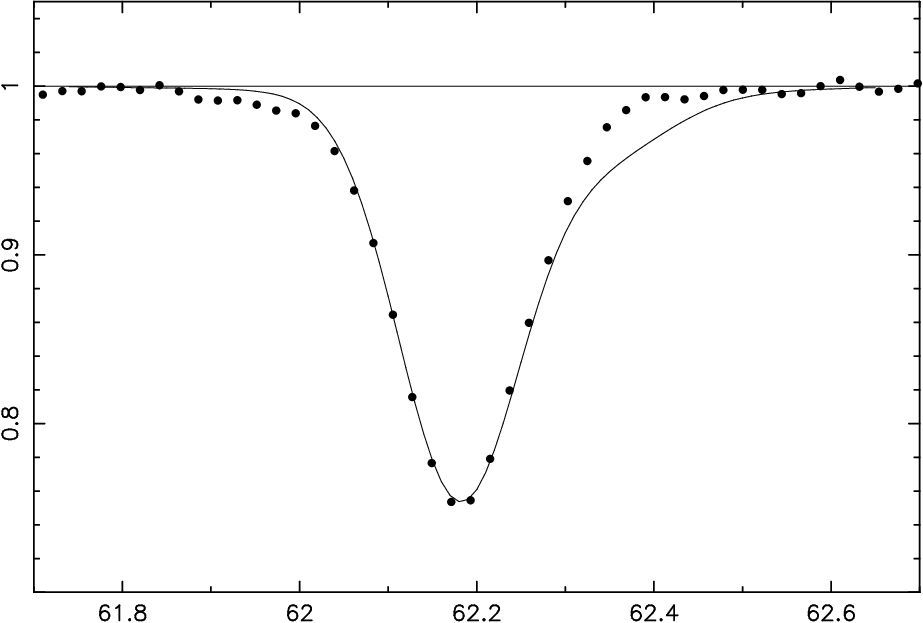}}
\caption{Comparison between the observed profile of the Ca I line at 
6162 \AA \ 
in HD 84937 and the computed profile if the $^6$Li feature would be 
due to 
a $^7$Li line of a companion star. The two profiles are clearly not 
compatible. }
\label{figure7}
\end{figure}

 Again, the binarity explanation of a fake $^6$Li is discarded. 
 Nevertheless we have asked to two groups
 to observe HD 84937 in speckle interferometry for checking for a 
possible companion.

\subsection{ Why is $^6$Li found, among  very metal-poor stars, 
 only in HD 84937 ?}
  
 It is very embarassing to have a single case of undisputable 
detection 
 of $^6$Li out of several very metal-poor stars  
   investigated  for the presence of $^6$Li. 
 What makes HD 84937 so special?
 We present in figure 8, the HR diagram of metal poor stars  
 with distances reliably established by Hipparcos. 
 Clearly, from Hipparcos parallaxes, HD 84937 is among the stars 
deficient 
 by more than a factor of 100 in metals, also  among the evolved 
subgiants,
 with BD+26 3578, and HD 140283. 
 It means that it is also  one of higher mass, less prone 
 to $^6$Li burning. HD 140283 is in principle
 even more favourable from this point of view, but its extremely low
 metallicity makes it  less good for having inherited enough $^6$Li 
by spallation, and
 also the star is possibly evolved enough to have diluted its $^6$Li 
 in a thick convective zone (Chaboyer \cite{Chaboyer94}). 
 Actually the  case the more similar to HD84937 is
 BD + 26 3578, for which $^6$Li detection has been claimed by 
 Smith  (\cite{Smith96})  and Smith et al. ( \cite{Smith98}). 

 We have theoretically investigated $^6$Li destruction al\-ong the ZAMS 
in metal-poor stars.
The details are described in another paper (Cayrel et al. 1999). 
The principle on which it is based is
that the burning rate has such a fast variation with temperature that 
$^6$Li , as well as
$^7$Li , at a given depth below the convective zone, either do not 
burn in 15 Gyr, or
burn so fast that they are destroyed in less than 1 Gyr, or so. The 
rate of decline of
the abundance of the element in the convective zone is then 
controlled by exchange of
matter between the convective zone and the deeper zone deprived very 
soon of the element.
This rate is a purely hydrodynamical quantity. So if for example, we 
find that $^7$Li
is strongly depleted in the convective zone when the burning layer is 
separated from
the convective zone by a layer having a thickness in mass equal to 
half of the mass 
of the convective zone, we shall assume that this will hold for 
$^6$Li also, taking
of course into account the fact that the location of the fast burning 
zone is moved
upwards for $^6$Li. This allows us to transform the $^7$Li 
depletion curve into a $^6$Li
depletion curve. Clearly, at metallicity [Fe/H]= -1.5, only the most 
massive stars
($M > 0.8 M_{\sun})$ can save their $^6$Li.
At [Fe/H]= -1., $^6$Li even burns {\it inside} the convective zone, so
there is no hope to find any $^6$Li.  We have not yet investigated 
the dilution effect,  above the 
turn off, but Chaboyer (\cite{Chaboyer94}) finds that this process 
becomes effective   at the position of HD 140283.
Fig. 8 shows the position of the stars in which $^6$Li has been found 
or is possibly present
with an abundance less than 10 per cent of the abundance of $^7$Li 
. They clearly cluster along
the turn off, in agreement with the theoretical expectation. These 
considerations
make understandable that $^6$Li can be found only in a very limited 
portion of the HR diagram. 

\begin{figure}
\resizebox{\hsize}{!}{\includegraphics{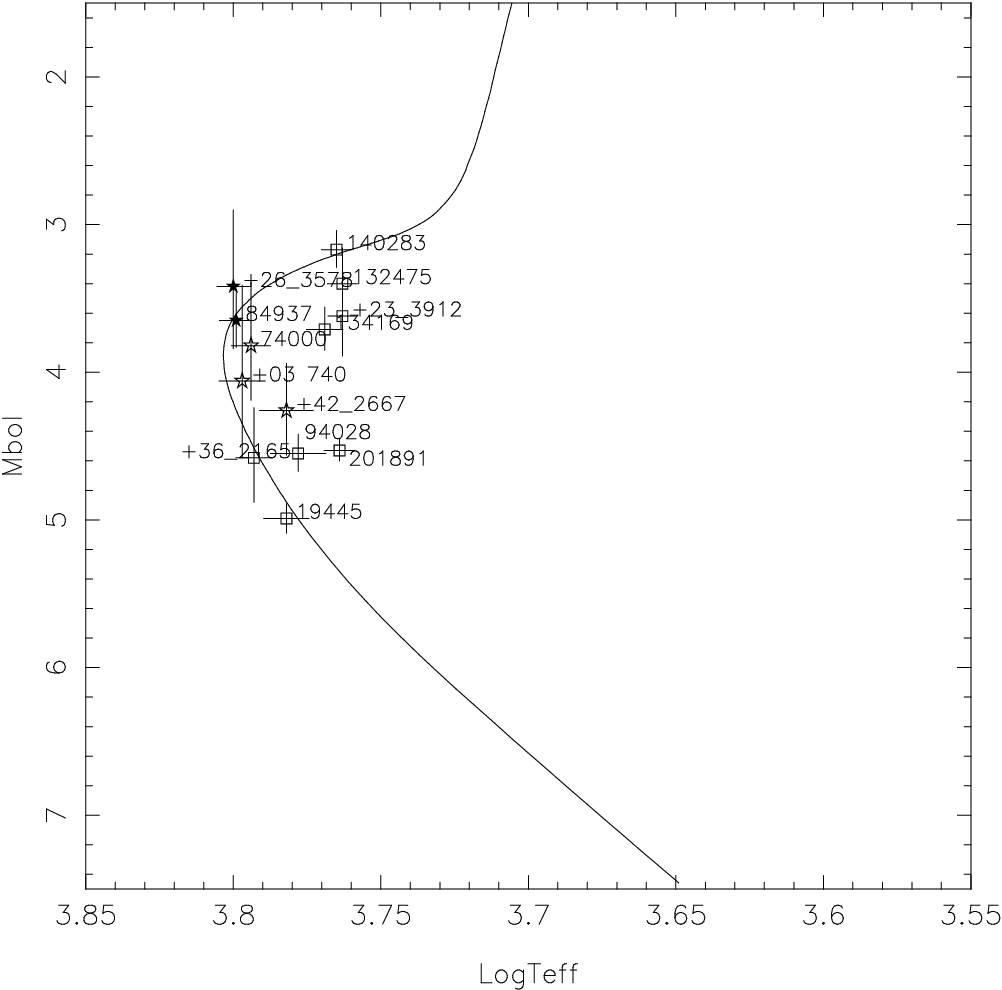}}
\caption{HR diagram of stars searched for $^6$Li. The symbols are: 
a filled star for an 
unambiguous measurement (only cases HD 84937 and BD+26 3578); an open star for a less 
solid 
claimed detection or  only an upper limit for $^6$Li, at a level of 10 to 13 per
cent of $^7$Li.
The squares are stars for which $^6$Li is undetectable, i.e. below 2 
per cent of $^6$Li.
 The curve is a 14 Gyr VandenBerg isochrone
for [Fe/H]= -2, shifted by 0.01 in log(Teff), as needed to fit the high 
parallax subdwarfs (Cayrel et al. 1997). It is interesting to note  
that the stars with detected, or possible, $^6$Li are fairly concentrated near 
the turnoff, suggesting that $^6$Li is burned below, and diluted above. All
stars have metallicities below [Fe/H]= -1.5, except BD+36 2165 and HD 201891 
which are near [Fe/H]= -1. In the name of the stars `HD' and `BD' have been 
omitted.}
\label{figure8}
\end{figure}
%
 
\section{The two other stars}

\begin{figure}
\resizebox{\hsize}{!}{\includegraphics{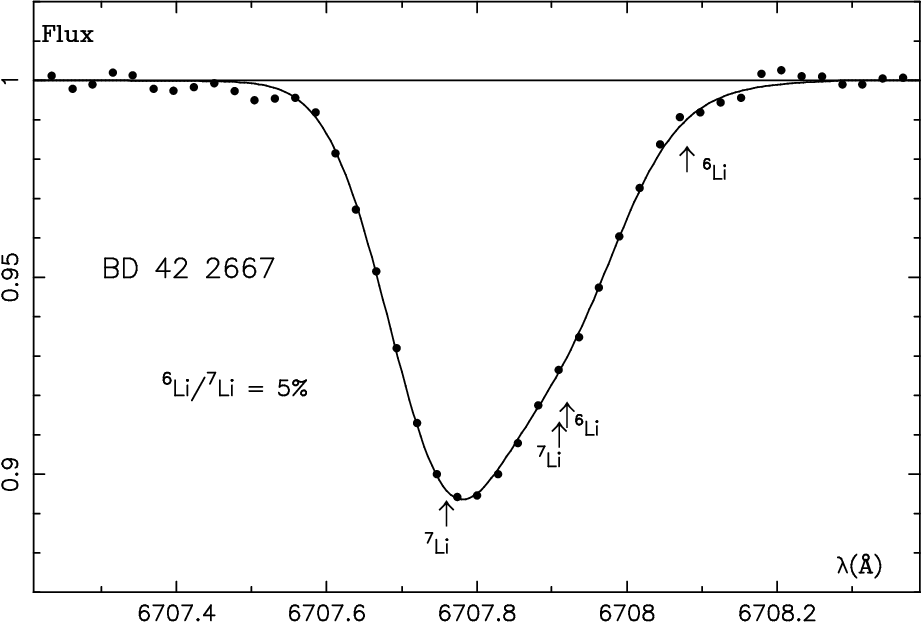}}
\caption{
 Best fit of the lithium blend for BD + 42 2667.  
The logaritmic abundances are respectively 2.088 
and 0.82 for
 $^7$Li and $^6$Li. A readjustment in wavelength of 5 m\AA \ was obtained.
 The FWHM is 0.125 and the rms of the fit 1.59E-03.      
  }
  \label{figure9}
 \end{figure}

\begin{figure}
 \resizebox{\hsize}{!}{\includegraphics{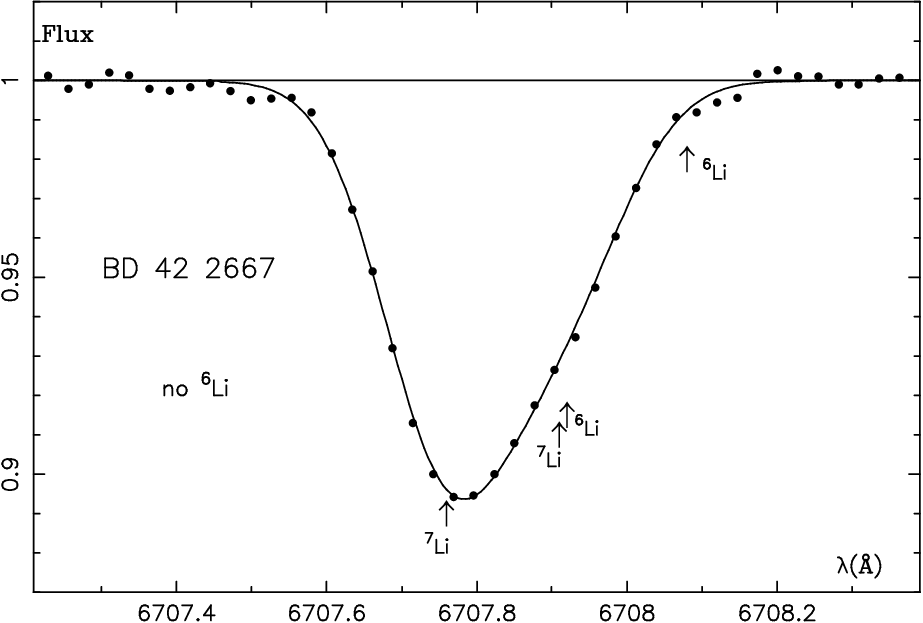}}
 \caption{
 Fit of the lithium blend for BD + 42 2667 with no $^6$Li.  
A wavelength shift of 0.008 \AA \  is needed.
  The logaritmic abundance of $^7$Li is  2.12 .
  The FWHM is 0.141 and the rms of the fit 1.75E-03.       
  }
  \label{figure10}
 \end{figure}
 
   Two other stars, BD +42 2667 and BD +36 2165 
  have been observed. The profile of the lithium line of the first 
star 
  is best explained by a $^6$Li / $^7$Li ratio of 10\%. And 
  the hypothesis of no $^6$Li provides a considerably worse fit. If 
again a 
  shift in wavelength is allowed, the profile may be explained by a
  $^6$Li / $^7$Li ratio of only 5\% (fig. 9), so that the value of 
the 
  abundance of $^6$Li may seem rather uncertain. 
   Assuming the 
  hypothesis of no $^6$Li, a not too bad fit may be found (fig.10), 
   
  These observations point towards a  detection of  $^6$Li in 
  this star, but the modest S/N ratio reached in the observation
  of this star precludes a conclusion at a very high confidence level.
  Smith et al. ( \cite{Smith98}) have also studied BD +42 2667 
  and conclude that it has
  probably no $^6$Li. We do not fight this conclusion, the residuals
  being unsufficiently apart in the two cases.

  The other observed star, \object {BD +36 2165}, does not provide 
  any definite detection of $^6$Li. The S/N ratio reached is even 
  lower, and does not enable any firm constraint.

\section{Discussion}

The observation of HD 84937 has  essential consequences, first on the
depletion of $^6$Li (and indirectly that of of $^7$Li) in Pop II 
stars,
and  on the cosmological status of Li/H observed in old
metal-poor stars, and also on $^6$Li (and Be and B) production.
The $^6$Li isotope is a pure spallation product (see Reeves 
\cite{Reev94} for a review).
This fragile nucleus is burnt at low temperature (about
$ 2.  10 ^6$ K) and cannot be synthesized inside stars. Spallation 
agents are
(i)  galactic cosmic rays (GCR), specifically acting in the galactic
disk through  p, $\alpha$ + He,CNO  $\rightarrow$  $^6$Li,$^7$Li and
(ii) in the halo phase ([Fe/H] $\leq  -1$), low energy $\alpha $,C and 
O nuclei ejected
and accelerated by supernovae, interacting with H and He
in the ISM (Cass\'e et al. \cite{Cas95}, Ramaty et al. \cite{Rama96}). 
This low
energy component (LEC) is likely to be responsible for the linear
relationship between Be, B and  [Fe/H]  discovered recently  
(Duncan et al. \cite{DPR97}, Molaro et al. \cite{MBCP97}, 
  Thorburn \& Hobbs \cite{TH96}, Primas  \cite{Pri96}
Garc\'{\i}a L\'opez et al.  \cite{Gar98}) as shown by
Vangioni-Flam et al. (\cite{VLC94}). This linear relationship,
specifically in the early Galaxy, is due to the fact that
freshly synthesized $\alpha $, C and O from SNe are accelerated at 
moderate energy
and fragment on the H, He nuclei in the ISM. Thus, the production rate
is independent of the ISM metallicity (which means that Be and B are 
`` primary''). 
The $^6$Li isotope itself
is also  produced by primary processes, through the two spallative 
processes 
GCR and LEC.
Consequently, its slope in the  ($\log $($^6$Li /H), [Fe/H]) plane is 
unity.

The $^6$Li abundance observed in the atmosphere of 
HD 84937 is a lower limit to that of the interstellar medium out of which
this star has formed since this isotope can be depleted in stellar
atmosphere due to proton capture and dilution with Li free
material coming from the interior. The depletion factor is still
uncertain in spite of many efforts to evaluate it (Lemoine et al. 
 \cite{lemoine},
Chaboyer  \cite{Chaboyer94}, Vauclair \& Charbonnel  \cite{Vau95}, 
Deliyannis \& Malaney 
\cite{Deliy95}).

If  $^6$Li is detected in the atmosphere of a Pop. II star, 
like HD 84937, one can infer reasonnably that the $^7$Li  
ab\-un\-dance is
essentially unaffected, since $^7$Li is destroyed at a higher 
stellar temperature than $^6$Li ,  and in this case the
Li/H ratio is close to the primordial one.

In this context, it is important to analyze $^6$Li data versus
$^7$Li and to follow the early rise of the $^6$Li/H ratio due to 
the 
$\alpha$, C, O + H,He processes described above, as a function of 
[Fe/H]
(evolutionary curve), since this theoretical $^6$Li/H value represents
at any time (metallicity) the initial abundance in the forming stars.
Since $^6$Li can only be depleted in stars, 
 the observed
points are located on this curve (no depletion) or below. Previous 
calculations (Vangioni-Flam et al.  \cite{VFCR97}) give 
the  Li/H versus [Fe/H] evolution (see their
figure 1)  where the spallative lithium 
is shown for LEC and GRC. This last component is weaker than the first.
Associated to the calculated $^6$Li / $^7$Li (their table 1) the 
LEC curve allows 
to predict the evolution of $^6$Li/H; at [Fe/H] = -2.3 corresponding 
to the metallicity of HD 84937, the theoretical $^6$Li/H is 
 slightly below the observation. This characterizes only a specific 
model.
 It is possible to find other scenarios able to reproduce all the 
LiBeB observational
 constraints, using metal-poor SN II compositions, well suited to the 
early
 Galaxy (Woosley \& Weaver \cite{WW95}). This kind of models meets the 
requirement
 of a high He/O source ratio, as pointed out by Smith et al. 
(\cite{Smith98}), 
 to explain the high $^6$Li /$^9$Be ratio. 
  This indicates that if $^6$Li is destroyed
and/or diluted, it is  only slightly . This is a challenge for the 
stellar
evolution theory. It is satisfying to see that 
in this case, the theoretical
description of the $^6$Li evolution in the early Galaxy is globally 
consistent with 
our observations. The occurrence of LEC  is strenghthened by this 
observation. This
theoretical model has some latitude depending on the composition
and energy spectrum of the beam adopted, as said above. It is 
possible
to find other source composition able to reproduce the LiBeB 
observational
constraints, including the uncomfortably high  $^6$Li / Be ratio found
here in HD 84937 by us, and other authors. A further discussion is 
postponed to 
a forthcoming paper  (Vangioni-Flam et al.  \cite{VFCCASS}) .\\
A word of caution, however. Our analysis is based on the abundance 
of $^6$Li
in one star, or two when the ratio $^6$Li~/~$^7$Li of 5 percent 
 in 
BD +26 3578 (Smith et al. \cite{Smith98}) will be confirmed. More precise 
Be abundances
in the two stars would be useful, as well as a richer sample of stars 
with measured $^6$Li.

Even assuming that the estimation of the initial $^6$Li
ab\-un\-dance is in error by a factor of 3, our observation shows 
that $^6$Li has
been depleted by  a factor of 3, at the most. The ratio of $^7$Li
depletion to $^6$Li depletion is model dependent. The burning of 
$^7$Li in
the region depleting $^6$Li by a factor of possibly 3 is close to 
negligible.
If $^7$Li is depleted,  it is in deeper layers, where we are sure 
that the 
mixing is slower, but by an amount which is very model dependent. 
Under
the simple model we have considered in section 4.3 the depletion of 
$^7$Li
cannot be more than 0.1 dex if the depletion of $^6$Li is below 0.5 
dex.

Regarding the cosmological consequences of this Li ab\-un\-dance
determination and analysis, it is interesting to note that the 
$^7$Li abundance $^7$Li/H = 1.6 10$^{-10}$ in HD 84937 :
 
(i) is for a star which belongs to the Spite plateau and  

(ii) is likely to 
represent the primordial value, once corrected for the small amount 
of $^7$Li 
produced with the $^6$Li by the LEC proces, since $^7$Li has not 
been affected 
by thermonuclear destructions (our above discussion). By comparing 
this Li/H value with the outcome of standard BBN calculations
(Sarkar  \cite{Sarkar97}, Schramm \& Turner  \cite{Schramm98}) 
this would lead to
$\eta _{10} \simeq $ 2.0 to 3.5, (or $\Omega_b \simeq $ 0.02 to 0.035 
for $H_0$= 60
km.s$^{-1}$.Mpc$^{-1}$), $Y_p \simeq$ 0.237 to 0.245, D$_p$/H $\simeq $ 
10$^{-4} $
to 5$\times $10$^{-5}$. This range on D$_p$/H is compatible
with both the high and low values of D/H measured in cosmological 
clouds on the
line of sight of quasars (Burles \& Tytler \cite{ BT98}, Webb et al. 
\cite{Webb97}).

More $^6$Li , $^7$Li observations in the atmospheres of old 
metal-poor stars could strengthen both the argument of occurence of 
the LEC process in the early history of the Galaxy, and the use of 
$^7$Li 
as a cosmological test, since the determination of the Spite 
plateau seems to reflect the true Big Bang value.  

\section{Conclusions}

 The main result of this paper is an analysis, with improved 
accuracy,   
 of the star HD 84937, leading to a new measurement of 
 of $^6$Li in HD 84937 .
 With a S/N of about 650 per pixel,(1020 on the filtered 
 spectrum)
 instead of  400 (Smith et al.  \cite{SLN93}) and  350 (Hobbs
 \& Thorburn  \cite{HT94})  reached previously, a null 
 abundance of $^6$Li is  now  more convincingly ruled out.
   
  The abundance found for
  $^6$Li ( 5.2 per cent of the abundance of $^7$Li) turns out to be 
  close the expected value of the 
  initial $^6$Li in HD 84937, using the evolutionary models 
  developed by Vangioni-Flam et al. (\cite{VFCR97}). 
 
  The process of formation of the LiBeB elements just quoted above
 is now 
shown to be  consistent also for $^6$Li, provided that the 
$\alpha+\alpha $ process is the dominant source of $^6$Li
in the early Galaxy.

The nearly equality between  the  predicted initial $^6$Li/H value 
and the observed value in HD 84937 shows that $^6$Li cannot have been 
strongly depleted in this star.  
 $^7$Li, more robust, 
 is necessarily still less  depleted, and its abundance can then 
 be taken as representative  of the cosmological primordial lithium.
 
\vspace{1cm}

\appendix{ \bf Appendix}
   
\vspace{0.5cm}
The convolution of the spectra by a narrow gaussian profile
of standard deviation $1/\sqrt{2}$  pixel, bringing an unsignificant
degradation in spectral resolution, but increasing the signal/noise
ratio by 57 per cent, make the classical $\chi^2 $ test inapplicable,
because the noise on  consecutive  pixels becomes correlated.
In order to  properly discuss the $(O-C)$ residuals, it is necessary 
to derive the expected mean value of the sum of the $((O-C)_i/\sigma_i)^2$ , 
and its
expected variance, in our specific case, where the independent
random variables are the noise on each data point, {\it before} 
 convolution. The convolved signal $Y_i$ is expressed by a linear
form of the unconvolved signal $y_i$ ({\it i} is the running number of 
the pixel along the Li feature):

$$ Y_i = \sum_{k=-2}^{2} w_k y_{i+k} $$

\begin{table}
\caption[]{Probability function of $X^2$ for 23 degrees of 
freedom \\ 
expectation value: 23, \hspace{0.2in} variance = 78.2 . \\
 The second column gives the probability to be in the bin ending at 
the entry 
 \\ The third column (cumulated values) are for the entry value
. }
\begin{tabular}{cccccc}
\noalign{\smallskip}
\hline
\noalign{\smallskip}
$X^2$ & prob. & cumul. & $X^2$ & prob. & cumul. \\
 1.0 & 0. & 0. &        26.0 & .0383 & .6818 \\
 2.0 & .0000 & .0000 & 27.0 &.0355 & .7172 \\
  3.0 &  .0000 &   .0000 & 28.0 & .0326 & .7498 \\
  4.0 & .0001 &   .0001 & 29.0  & .0296 & .7794\\
  5.0 & .0004 & .0005 & 30.0 & .0268 & .8062 \\
  6.0 & .0011 & .0016 & 31.0 & .0242 & .8304 \\
  7.0 & .0028 & .0044 & 32.0 & .0216 & .8521 \\
  8.0 & .0052 & .0096 & 33.0 & .0190 & .8711 \\
  9.0 & .0088 & .0184 & 34.0 & .0171 & .8881 \\
 10.0 & .0133 & .0318 & 35.0 & .0150 & .9032 \\
 11.0 & .0187 & .0505 & 36.0 & .0133 & .9165 \\
 12.0 & .0239 & .0744 & 37.0 & .0116 & .9281 \\
 13.0 & .0298 & .1042 & 38.0 & .0100 & .9381 \\
 14.0 & .0351 & .1392 & 39.0 & .0088 & .9470 \\
 15.0 & .0401 & .1793 & 40.0 & .0076 & .9546 \\ 
 16.0 & .0436 & .2230 & 41.0 & .0065 & .9611 \\
 17.0 & .0462 & .2692 & 42.0 & .0057 & .9668 \\
 18.0 & .0482 & .3174 & 43.0 & .0049 & .9717 \\
 19.0 & .0496 & .3669 & 44.0 & .0043 & .9760 \\
 20.0 & .0494 & .4164 & 45.0 & .0036 & .9796 \\
 21.0 & .0492 & .4656 & 46.0 & .0031 & .9827 \\
 22.0 & .0478 & .5134 & 47.0 & .0027 & .9855 \\
 23.0 & .0455 & .5589 & 48.0 & .0022 & .9876 \\
 24.0 & .0435 & .6024 & 49.0 & .0020 & .9897 \\
 25.0 & .0410 & .6435 & 50.0 & .0016 & .9913 \\
 
 \noalign{\smallskip}
\hline
\end{tabular}

\end{table}
 
with, in our particular case, the numerical values:

0.01, 0.208, 0.564, 0.208, 0.01 for $w_{-2 }$ to $w_{2}$ , 
respectively.

The pseudo-$\chi^2 $ , let us call it $X^2$:

$$ X^2 =\sum_{i=i_1}^{i_2} ({Y_i-C_i \over \sigma ^{\prime }_i})^2 $$

 can be readily computed ($i_1$ and $i_2$ are respectively the first 
 and the last pixel of the lithium feature),
 assuming that the $y_i$ are affected by a gaussian noise equal to the
 noise on each pixel value of the spectrum before convolution ( 1/650 
 in fraction of the continuum level ). It is a well known result
 (linear combination of gaussian variables) that  the mathematical 
 expectation of
$X^2$ is reduced with respect to the mathematical expectance of
 the true $\chi^2$ in the ratio: 

 $$ r= \sum_{k=-2}^{2} w_k^2 = 0.4048 $$

This gives the value of $\sigma^\prime$

$$ \sigma^\prime = \sqrt{r}= 0.636\sigma $$

The computation of the complete distribution has been performed by 
running
1~000~000 realizations, and binning the resulting values of $X^2$. 
The results are given
in table 1 of this  appendix for 23 degrees of freedom. This 
computation
shows that the variance is 3.4 times the mathematical expectance, 
instead of two 
 for a true $\chi^2$ law. So, part of the gain in noise reduction is 
 lost in
 a shallower distribution, but the gain is anyhow substantial, as 
 already shown, in section 4.1, where the two $\chi^2$ were used 
 competitively.
 Several simulations, done on test cases, have confirmed 
 the effectiveness
 of noise filtering, with the use of a modified $\chi^2$ distribution.  

\begin{acknowledgements} F. and M. Spite would like to acknowledge 
useful discussions with S. Vauclair and C. Charbonnel, and a grant of 
the Programme National de Cosmologie which enabled the data 
processing and the computations made for this work. We are pleased 
to acknowledge the use of the SIMBAD database at the CDS, 
including the use of the HIP catalogue. We are indebted to Michel 
Mayor for providing useful information about the radial velocity of 
HD 84937 in advance of publication. We thank D. Pelat for a useful clarification
in the computation of the ``standard'' error in a $\chi^2$ test.
 The authors are grateful to C. Sneden
for fruitful discussions and for improving the language of the paper,
to P.E.  Nissen for making available the preprint of his last paper
with Smith and Lambert, to Martin Asplund for communicating his
preliminary results
on blue shifts computed from hydrodynamical simulations, and to P. Molaro for
important remarks to improve several sections of our paper. 

\end{acknowledgements}

\end{document}